\documentclass[a4paper,12pt]{article}

\pdfoutput=1

\usepackage{amsmath}
\usepackage{amssymb}
\usepackage{mathrsfs}
\usepackage{bbm}
\usepackage{graphicx,subfigure}
\usepackage[numbers,sort&compress]{natbib}

\usepackage{color}
\usepackage{ulem}

\numberwithin{equation}{section}

\newlength{\dinwidth}
\newlength{\dinmargin}
\def\GeV{\,{\rm GeV}}
\def\TeV{\,{\rm TeV}}
\def\SM{{\rm SM}}
\def\THDM{{\rm 2HDM}}
\def\Unitary{\rm Unitary}
\def\B{\overline{\mathcal B}}

\def\xHC{x_{H^\pm}}
\def\xt{x_t}

\makeatletter
\newcommand{\thickhline}{%
    \noalign {\ifnum 0=`}\fi \hrule height 1pt
    \futurelet \reserved@a \@xhline
}
\makeatother

\begin{document}

\title{\bf \Large Revisiting $B_s \to \mu^+\mu^-$ in the two-Higgs doublet
  models \\[-0.6em] with $Z_2$ symmetry}

\author{
Xiao-Dong Cheng$^{a,b}$\footnote{chengxd@mails.ccnu.edu.cn},
Ya-Dong Yang$^a$\footnote{yangyd@mail.ccnu.edu.cn}\,
and
Xing-Bo Yuan$^{c,d}$\footnote{xbyuan@yonsei.ac.kr}}


\date{}
\maketitle

\vspace{-3.6em}

\begin{center}
\textit{$^a$Institute of Particle Physics and Key Laboratory of Quark and Lepton Physics (MOE), \\[-0.6em] Central China Normal University, Wuhan, Hubei 430079, P. R. China}\\[0.5em]
\textit{$^b$College of Physics and Electronic Engineering, Xinyang Normal University, \\[-0.6em]Xinyang, Henan 464000, P. R. China}\\[0.5em]
\textit{$^c$Department of Physics and IPAP, Yonsei University, Seoul 120-749, Korea}\\[0.5em]
\textit{$^d$State Key Laboratory of Theoretical Physics, Institute of Theoretical Physics, \\[-0.6em] Chinese Academy of Sciences, Beijing 100190, P. R. China}
\end{center}

\vspace{1em}

\begin{abstract}
{\noindent}We revisit the rare leptonic decay $B_s \to \mu^+ \mu^-$ in the two-Higgs doublet models with a softly broken $Z_2$ symmetry, namely type-I, type-II, type-X and type-Y 2HDMs. We have derived the relevant full one-loop Wilson coefficients of the four 2HDMs from the recent calculation in the aligned two-Higgs doublet model by Li, Lu and Pich, which could  be mapped to all the four 2HDMs for both large and small $\tan\beta$. It is found that a new term associated with the soft $Z_2$ symmetry breaking parameter $M$ can be enhanced by $\tan^2\beta$ in the type-II 2HDM, which has not been considered in the literature. 
 Imposing both theoretical and experimental constraints, we have renewed the bounds on the  parameter spaces of the four 2HDMs.  Different from our previous paper, however, we find that all the four 2HDMs give sizable and similar contributions to $\B(B_s \to \mu^+ \mu^-)$ within the stringently restricted parameter spaces,  but very tiny as regards the mass-eigenstate rate asymmetry $\mathcal A_{\Delta\Gamma}$; this makes it unfeasible to discriminate the four types of 2HDM with the correlations between the observables in  $B_s \to \mu^+ \mu^-$ decay.
\end{abstract}

\newpage

\section{Introduction}

The discovery~\cite{ATLAS:12,CMS:12} of a new boson with a mass close to 125 GeV has been well anticipated as the standard model Higgs boson~\cite{ATLAS:13,CMS:13,LHC:15} and provided the first experimental evidence of the Higgs mechanism~\cite{Englert:64,Higgs:64,Guralnik:64}. It is a great triumph, but not an end, of the giant campaign for Higgs hunting in the development of particle physics. Although the subsequent more precise measurements~\cite{ATLAS:14a,ATLAS:14b,ATLAS:15,CMS:14a,CMS:14b} at the LHC have shown the properties of the Higgs boson are well consistent with the predictions of the standard model (SM), the precision of the current experimental data still leave open the possibility of an extended Higgs sector~\cite{Giardino:13,Belanger:13}. Among many new physics scenarios beyond the SM, the two Higgs doublet models (2HDM)~\cite{Lee:73,Gunion:89,Branco:11} are the simplest extensions of the SM.


In the 2HDMs, an additional Higgs doublet is introduced to the SM Higgs sector, which could result in rich phenomena, in collider physics~\cite{Kaffas:07,Aoki:09,Arhrib:09,Ko:11,Bai:12,Chen:13a,Ko:13,Celis:13,Jung:15,Abbas:15}, flavor physics~\cite{Xiao:06,Chang:08,Mahmoudi:09,Crivellin:12,Hermann:12,Celis:12,Li:13,Kim:15}, neutrino physics~\cite{Kanemura:13}, dark matter~\cite{Ma:06,Barbieri:06,Ko:12} and cosmology~\cite{Shu:13,Dorsch:13}. However, unlike the SM, unwanted tree-level flavor-changing neutral current (FCNC) interactions in the 2HDM are not forbidden by the Glashow-Illiopoulos-Maiani (GIM) mechanism. Besides some other solutions~\cite{Branco:96,Buras:00,D'Ambrosio:02,Manohar:06,Pich:09}, this issue is usually addressed by the Natural Flavor Conservation (NFC) hypothesis through imposing a discrete $Z_2$ symmetry~\cite{Glashow:76}. According to different $Z_2$ charge assignments, there are four types of the NFC 2HDM, referred to as the type-I, type-II, type-X and type-Y 2HDM, respectively. Of course, there are new parameters in the 2HDMs to be determined or excluded by the measurements of electro-weak processes. To this end, B-meson decays are usually employed to constrain their parameter spaces.

Among the rare B-meson decays, the leptonic processes $B_q \to \mu^+ \mu^-$ ($q=d$, or $s$) are of special interest~\cite{Buras:12,Arbey:12}. They suffer from very few hadronic uncertainties and are induced by FCNC transitions, which make them sensitive probes to the effects of physics beyond the SM, especially models with a non-standard Higgs sector~\cite{He:88,Savage:91,Chankowski:00,Dedes:08,Alok:08}. Recently, the next-to-leading order (NLO) electroweak corrections and the next-to-next-to-leading order (NNLO) QCD corrections~\cite{Bobeth:13a,Bobeth:13b,Hermann:13} in the SM have been calculated. On the BSM side, a full one-loop calculation in the aligned 2HDM (A2HDM) has been performed in ref.~\cite{Li:14}.

Motivated by this progress, in this paper we perform a detailed study of the $B_s \to \mu^+ \mu^-$ decay within the 2HDMs with $Z_2$ symmetry. At present, this process is calculated in the type-II 2HDM in large $\tan\beta$ limit only~\cite{Logan:00,Huang:00,Bobeth:01}. Using the Higgs base correspondence between the A2HDM and the 2HDMs, we will derive the relevant full one-loop Wilson coefficients of the four variant 2HDMs contributing to the $B_s \to \mu^+ \mu^-$ decay from the recent A2HDM results~\cite{Li:14} without the large $\tan\beta$ approximation. We also investigate the possibility to discriminate the four different types of 2HDM in the light of the recent collider and flavor physics data, as an update of our previous work~\cite{Cheng:14}. We combine the constraints from $B_{s,d}\to \mu^+ \mu^-$, $B_{s,d}-\bar B_{s,d}$ mixing, $B\to \tau \nu$ and $\bar B \to X_s \gamma$~\cite{Huber:14,Misiak:15}, with the experimental data from the direct search for Higgs bosons at LEP~\cite{ALEPH:02}, Tevatron~\cite{CDF:05,D0:09} and LHC~\cite{CMS:12a,ATLAS:13a}, and the constraints from perturbativity, tree-level vacuum stability and perturbative unitary. For the $B_s \to \mu^+ \mu^-$ decay, the correlations between its branching ratio and the mass-eigenstate rate asymmetry $\mathcal A_{\Delta\Gamma}$ are also reevaluated with the constrained parameter space of the 2HDMs obtained in this paper. We have found that $\mathcal A_{\Delta\Gamma}$  can slightly deviate from the SM prediction in the type-II 2HDM only, and that the ratio of time-integrated $\B(B_s \to \mu^+ \mu^-)$ gets similar contributions from the four 2HDMs; this makes it very hard to discriminate the four types of 2HDMs with the correlation between $\mathcal A_{\Delta\Gamma}$ and $\B(B_s \to \mu^+ \mu^-)$ as suggested in our previous work~\cite{Cheng:14}.

The paper is organized  as follows. In section~\ref{sec:Bs_lm_lm}, we give a brief overview of the $B_s \to \mu^+ \mu^-$ decay. In section~\ref{sec:WC}, full one-loop contributions from the 2HDMs with $Z_2$ symmetry are derived explicitly. In section~\ref{sec:analysis}, we give our detailed numerical results and discussions. We conclude in section~\ref{sec:conclusion}. The relevant theoretical formulas are recapitulated in the Appendix.

\section{$B_s \to \mu^+ \mu^-$ in the SM}\label{sec:Bs_lm_lm}
In the SM, the leptonic decays $B_q \to \mu^+ \mu^-$ ($q=d$ or $s$) arise from the $W$ box and $Z$ penguin diagrams. Generally, these decays can be described by the low-energy effective Hamiltonian
\begin{align}\label{eq:Hamiltonian}
  \mathcal H_{\rm eff}=-\frac{G_F}{\sqrt 2}\frac{\alpha_e}{\pi s_W^2}V_{tb} V_{tq}^* \bigl( C_{10} \mathcal O_{10} + C_S \mathcal O_S + C_P \mathcal O_P \bigr),
\end{align}
where $\alpha_e$ denotes the QED fine-structure constant and $V_{ij}$ the CKM matrix elements. The semi-leptonic operators are defined as
\begin{align}
  \mathcal O_{10}=\bigl ( \bar q \gamma_\mu P_L b \bigr) \bigl( \bar \mu \gamma^\mu \gamma_5 \mu \bigr),\quad
  \mathcal O_S=\frac{m_\mu m_b}{m_W^2} \bigl( \bar q P_R b \bigr) \bigl( \bar\mu \mu \bigr) ,\quad
  \mathcal O_P=\frac{m_\mu m_b}{m_W^2}  \bigl(\bar q P_R b \bigr) \bigl( \bar \mu \gamma_5 \mu \bigr).
\end{align}
In the SM, the contributions from the scalar operators $\mathcal O_S$ and $\mathcal O_P$ are highly suppressed (the corresponding Wilson coefficients are given in eq.~\eqref{eq:WCSPSM}.),  but $C_{10}$ will play the dominant role. Its explicit expressions up to the NLO QCD corrections can be found in ref.~\cite{Buchalla:93,Misiak:99,Buchalla:98}. Recently, calculations of the NLO EW~\cite{Bobeth:13b} and NNLO QCD~\cite{Hermann:13} corrections have also been completed~\cite{Bobeth:13a}. This progress will be incorporated into our calculations.

With the effective Hamiltonian eq.~\eqref{eq:Hamiltonian}, the branching ratio of $B_q \to \mu^+ \mu^-$ reads
\begin{align}
  \mathcal B(B_q\to \mu^+\mu^-)=\frac{\tau_{B_q}G_F^4
    m_W^4}{8\pi^5}|V_{tb}V_{tq}^*|^2f_{B_q}^2m_{B_q}m_\mu^2\sqrt{1-\frac{4
    m_\mu^2}{m_{B_q}^2}}\bigl(|P|^2+|S|^2\bigr),
\end{align}
where $m_{B_q}$, $\tau_{B_q}$ and $f_{B_q}$ denote the mass, mean lifetime and decay constant of $B_q$ meson respectively. The short-distance contributions $S$ and $P$ are defined as
\begin{align}\label{eq:SP}
  P=C_{10}+\frac{m_{B_q}^2}{2m_W^2}\left(\frac{m_b}{m_b+m_q}\right)C_P,\qquad
  S= \sqrt{1-\frac{4m_\mu^2}{m_{B_q}^2}}\frac{m_{B_q}^2}{2m_W^2}\left(\frac{m_b}{m_b+m_q}\right)C_S.
\end{align}
As discussed in the following  section, there is no BSM phase in the 2HDMs with $Z_2$ symmetry. Therefore, we only consider the case that both $S$ and $P$ are real in this paper.

As pointed out in ref.~\cite{DeBruyn:12}, the measured branching ratio of $B_q \to \mu^+ \mu^-$ should be the time-integrated one, denoted by $\B (B_q \to \mu^+ \mu^-)$. In order to compare with the experimental measurements, the sizable effect of $B_s - \bar B_s$ oscillations should be taken into account~\cite{DeBruyn:12,Genon:11}, and one has
\begin{align}
  \B (B_s\to \mu^+\mu^-)&=\left(\frac{1+\mathcal A_{\Delta\Gamma}y_s}{1-y_s^2}\right)\mathcal   B(B_s\to \mu^+\mu^-),\nonumber\\
  \B (B_d\to \mu^+\mu^-)&\approx\mathcal  B(B_d\to \mu^+\mu^-),
\end{align}
where the mass-eigenstate rate asymmetry $\mathcal A_{\Delta\Gamma}$ can be expressed as
\begin{align}\label{eq:A}
  \mathcal A_{\Delta\Gamma}=\frac{|P|^2-|S|^2}{|P|^2+|S|^2}.
\end{align}
The observable $\mathcal A_{\Delta\Gamma}$ is independent of the branching ratio of $B_s \to \mu^+ \mu^-$ and provides complementary information on the short-distance structure of this decay. In the SM, $\mathcal A_{\Delta\Gamma}=+1$.

Following ref.~\cite{DeBruyn:12}, it is convenient to introduce the ratio
\begin{align}\label{eq:R}
  R\equiv \frac{\overline{\mathcal B}(B_s\to \mu^+ \mu^-)}{\mathcal
    B(B_s\to \mu^+ \mu^-)_{\rm SM}}=\left(\frac{|P|^2}{1-y_s}+\frac{|S|^2}{1+y_s}\right)\frac{1}{|S_{\rm SM}|^2+|P_{\rm SM}|^2},
\end{align}
where both hadronic uncertainties and CKM matrix elements are canceled out.

\section{$B_s \to \mu^+ \mu^-$ in the 2HDMs with $Z_2$ symmetry}\label{sec:WC}
In the 2HDMs with $Z_2$ symmetry, $b \to s \mu^+ \mu^-$ processes receive contributions from box diagrams with charged Higgs and penguin diagrams with $Z$ boson and neutral Higgs bosons. The Wilson coefficient $C_{10}$ has been calculated in the type-II 2HDM~\cite{Chankowski:00}. For $C_S$ and $C_P$, only the leading contributions in the large $\tan\beta$ limit have been computed in the type-II model~\cite{Logan:00,Huang:00,Bobeth:01}. However, the remaining contributions could be important for some specific $\tan\beta$ values in the other types of 2HDMs. In this section, we first of all give a brief introduction to the 2HDMs with $Z_2$ symmetry, and then show that the Wilson coefficients could be derived explicitly from the recent full one-loop results of the A2HDM~\cite{Li:14}. 

\subsection{2HDMs with $Z_2$ symmetry}\label{sec:2HDM}
The 2HDM extends the SM Higgs sector with an additional scalar doublet. With the two Higgs doublets $\Phi_1$ and $\Phi_2$, the CP-conversing 2HDM potential with a softly broken $Z_2$ symmetry reads~\cite{Branco:11}
\begin{align}\label{eq:HiggsPotential}
V=&+m_1^2\Phi_1^\dagger\Phi_1
+m_2^2\Phi_2^\dagger\Phi_2-m_3^2\bigl(\Phi_1^\dagger\Phi_2+\Phi_2^\dagger\Phi_1\bigr)\nonumber\\
&+\frac{\lambda_1}{2}\bigl(\Phi_1^\dagger\Phi_1\bigr)^2+\frac{\lambda_2}{2}\bigl(\Phi_2^\dagger\Phi_2\bigr)^2+\lambda_3\bigl(\Phi_1^\dagger\Phi_1\bigr)\bigl(\Phi_2^\dagger\Phi_2\bigr)+\lambda_4\bigl(\Phi_1^\dagger\Phi_2\bigr)\bigl(\Phi_2^\dagger\Phi_1\bigr)\nonumber\\
&+\frac{\lambda_5}{2}\bigl[\bigl(\Phi_1^\dagger\Phi_2\bigr)^2+\bigl(\Phi_2^\dagger\Phi_1\bigr)^2\bigr],
\end{align}
where $m_3^2(\Phi_1^\dagger \Phi_2 + \Phi_2^\dagger \Phi_1)$ is a soft $Z_2$ symmetry breaking term and the parameters $m_{1-3}$ and $\lambda_{1-5}$ are real. The two Higgs doublets $\Phi_1$ and $\Phi_2$ can be generally parameterized as
\begin{align}
\Phi_i=
\begin{pmatrix}
\omega_i^+\\
\frac1{\sqrt2}(v_i+h_i-iz_i)
\end{pmatrix},
\end{align}
where the two vacuum expectation values ($vev$) $v_1$ and $v_2$ are real and positive. From the vacuum condition~\cite{Kanemura:04}
\begin{align}\label{eq:vac}
m_3^2 v_2 -m_1^2 v_1 -\frac{1}{2} \lambda_1 v_1^3 -\frac{1}{2}\lambda_{345} v_1 v_2^2&=0, \nonumber\\
m_3^2 v_1 -m_2^2 v_2 -\frac{1}{2} \lambda_2 v_2^3 -\frac{1}{2}\lambda_{345} v_1^2 v_2 &=0,
\end{align}
they can be expressed as other parameters in the Higgs potential, where $\lambda_{345}=\lambda_3+\lambda_4+\lambda_5$ is defined. By introducing the $vev$ $v$  ($v=v_{\rm SM}=246\GeV$), the mixing angle $\beta$ and the soft $Z_2$ symmetry breaking parameter $M$ as $v_1=v\cos\beta$, $v_2=v\sin\beta$ and $M^2=m_3^2/s_\beta c_\beta$, we can use $(v,\beta,M,\lambda_{1-5})$ as independent 2HDM potential parameters.

Physical Higgs states are obtained by the following rotations:
\begin{align}
\begin{pmatrix}h_1\\h_2\end{pmatrix}=R(\alpha)
\begin{pmatrix}H\\h\end{pmatrix},\quad
\begin{pmatrix}z_1\\z_2\end{pmatrix}=R(\beta)
\begin{pmatrix}G^0\\A\end{pmatrix},\quad
\begin{pmatrix}\omega_1^+\\\omega_2^+\end{pmatrix}=R(\beta)
\begin{pmatrix}G^+\\H^+\end{pmatrix},
\end{align}
where the rotation matrix is given by
\begin{align}
R(\theta)=\begin{pmatrix}\cos\theta&-\sin\theta\\
\sin\theta&\cos\theta\end{pmatrix}.
\end{align}
The mixing angle $\alpha$ is determined by the Higgs potential of eq.~\eqref{eq:HiggsPotential}~\cite{Kanemura:04},
\begin{align}\label{eq:alpha}
  \tan2\alpha&=\frac{(M^2-\lambda_{345}v^2)s_{2\beta}}{(M^2-\lambda_1 v^2)c_\beta^2 - (M^2-\lambda_2 v^2)s_\beta^2}.
\end{align}
In the 2HDM with $Z_2$ symmetry, the physical Higgs spectrum consists of five degrees of freedom: two charged scalars $H^\pm$, two CP-even neutral scalars $h$ and $H$, and one CP-odd neutral scalar $A$. The quartic couplings $\lambda_i$ in the Higgs potential can be expressed in terms of their masses as~\cite{Kanemura:04}
\begin{align}\label{eq:lam:1}
\lambda_1&=\frac{1}{v^2 c_\beta^2} \left(-s_\beta^2 M^2 + s_\alpha^2 m_h^2 + c_\alpha^2 m_H^2 \right) ,   \nonumber\\
\lambda_2&=\frac{1}{v^2 s_\beta^2} \left(-c_\beta^2 M^2 + c_\alpha^2 m_h^2 + s_\alpha^2 m_H^2 \right) ,   \nonumber\\
\lambda_3&=-\frac{M^2}{v^2} +2\frac{m_{H^\pm}^2}{v^2} +\frac{1}{v^2}\frac{s_{2\alpha}}{s_{2\beta}} \left(m_H^2-m_h^2\right),\nonumber\\
\lambda_4&=\frac{1}{v^2} \left(M^2 + m_A^2 -2 m_{H^\pm}^2\right), \nonumber\\
\lambda_5&=\frac{1}{v^2} \left(M^2-m_A^2\right).
\end{align}
Therefore, the eight parameters in the Higgs potential $m_{1-3}$ and $\lambda_{1-5}$ can be rewritten equivalently by the four physical Higgs masses $m_h$, $m_H$, $m_A$, $m_{H^\pm}$, the two mixing angles $\alpha$ and $\beta$, the $vev$ $v=v_{\rm SM}$, and the $Z_2$ symmetry breaking parameter $M$. In the case of $\lambda_1=\lambda_2$, which is considered in ref~\cite{Huang:00,Bobeth:01}, $M$ can be eliminated and the 2HDM potential parameters can be expressed by seven parameters $(\alpha,\beta,v,m_h, m_H, m_A,m_{H^\pm})$ as
\begin{align}\label{eq:lam:2}
  \lambda_1=\lambda_2&=\frac{1}{2v^2}(m_h^2+m_H^2)-\frac{1}{2v^2} \frac{c_{2\alpha}}{c_{2\beta}} (m_h^2-m_H^2),\nonumber\\
  \lambda_3&=-\frac{1}{2v^2}(m_h^2+m_H^2-4m_{H^\pm}^2)-\frac{1}{2v^2}(m_h^2-m_H^2)\left(\frac{c_{2\alpha}}{c_{2\beta}}+2\frac{s_{2\alpha}}{s_{2\beta}}\right),\nonumber\\
  \lambda_4&=\frac{1}{v^2}(m_A^2-2m_{H^\pm}^2)+\frac{1}{2v^2}(m_h^2+m_H^2)+\frac{1}{2v^2}\frac{c_{2\alpha}}{c_{2\beta}}(m_h^2-m_H^2),\qquad\qquad\qquad {\rm for}\,\,\, \lambda_1=\lambda_2.\nonumber\\
  \lambda_5&=-\frac{m_A^2}{v^2}+\frac{1}{2v^2}(m_h^2+m_H^2) +\frac{1}{2v^2}\frac{c_{2\alpha}}{c_{2\beta}}(m_h^2-m_H^2),\nonumber\\
  M^2&=\frac{1}{2}(m_h^2+m_H^2) + \frac{1}{2}\frac{c_{2\alpha}}{c_{2\beta}}(m_h^2-m_H^2),
\end{align}

In the interaction basis, the general Yukawa Lagrangian of the 2HDM can be written as
\begin{align}
- \mathcal L_Y=
\bar Q_L( Y_1^d\Phi_1+Y_2^d\Phi_2) d_R
+\bar Q_L(Y_1^u \tilde\Phi_1+Y_2^u \tilde\Phi_2)u_R
+\bar L_L(Y_1^\ell\Phi_1+Y_2^\ell\Phi_2) e_R
+\text{H.c.},
\end{align}
where $\tilde\Phi_i=i\sigma_2 \Phi_i^*$, $Q_L$ and $L_L$ denote the SM quark and lepton doublets, and $u_R$, $d_R$, and $e_R$ are the right-handed up-type quark, down-type quark and lepton singlet, respectively. The Yukawa coupling matrices $Y_i^{u,d,\ell}$ are $3\times 3$ complex matrices in flavor space.

\begin{table}[t]
\centering
\begin{tabular}{l c c c c c c }
\hline 
             & $\Phi_1$ & $\Phi_2$ & $u_R$ & $d_R$ & $\ell_R$ & $Q_L$, $L_L$ \\  \thickhline
Type-I  & $+$ & $-$ & $-$ & $-$ & $-$ & $+$ \\
Type-II & $+$ & $-$ & $-$ & $+$ & $+$ & $+$ \\
Type-X  & $+$ & $-$ & $-$ & $-$ & $+$ & $+$ \\
Type-Y  & $+$ & $-$ & $-$ & $+$ & $-$ & $+$ \\
\hline
\end{tabular}
\caption{\small Charge assignments of the $Z_2$ symmetry in the four types of 2HDM.} 
\label{tab:type}
\end{table}

In order to avoid tree-level FCNC, a discrete $Z_2$ symmetry is introduced~\cite{Glashow:76}. All the possible nontrivial $Z_2$ charge assignments are listed in table~\ref{tab:type}, which define the four well-known types of 2HDM, i.e. type-I, type-II, type-X and type-Y. In the mass-eigenstate basis, the Yukawa interactions can be written in the form
\begin{align}
-{\mathcal L}_Y=&+\sum_{f=u,d,\ell} \left[m_f \bar f f+\left(\frac{m_f}{v}\xi_h^f \bar f fh+\frac{m_f}{v}\xi_H^f \bar f fH-i\frac{m_f}{v}\xi_A^f \bar f \gamma_5fA \right) \right]\nonumber\\
&+\frac{\sqrt 2}{v}\bar u \left (m_u V \xi_A^u P_L+ V m_d\xi_A^d P_R \right )d H^+ +\frac{\sqrt2m_\ell\xi_A^\ell}{v}\bar\nu_L \ell_R H^+
+\text{H.c.},\label{Eq:Yukawa}
\end{align}
where $P_{L,R}=(1\mp \gamma_5)/2$. The Yukawa couplings $\xi_{h,H,A}^f$ in the four types of 2HDM are listed in table~\ref{tab:Yuk}. In addition, the couplings of the light CP-even Higgs boson $h$ to gauge bosons $W^+W^-$ or $ZZ$ can be written as $g_{hVV}=\sin(\beta-\alpha)g_{hVV}^{\rm SM}$, which is normalized to the corresponding couplings of the SM Higgs boson $g_{hVV}^{\rm SM}$~\cite{Branco:11}.

Recently, the LHC Run I data confirm the SM Higgs-like nature of the $125\GeV$ boson discovered at the LHC~\cite{ATLAS:13,CMS:13,LHC:15}. If the light CP-even Higgs $h$ in the 2HDM is identified with the observed $125\GeV$ boson, global fits to the LHC Higgs data suggest that all four types of 2HDM should lie close to the so-called alignment limit~\cite{Coleppa:13,Chen:13,Eberhardt:13,Dumont:14a,Dumont:14b,Bernon:14,Chowdhury:15}
\begin{align}
    \sin(\beta-\alpha)=1,
\end{align}
where both the Yukawa and the gauge couplings of $h$ are identical to the values of the SM Higgs boson. From eqs.~\eqref{eq:vac} and \eqref{eq:alpha}, the alignment limit can be achieved when the quartic couplings in the Higgs potential satisfy \cite{Carena:2013ooa,Carena:2014nza,Dev:2014yca}
\begin{align}
  \tan^2\beta=\frac{\lambda_1-\lambda_{345}}{\lambda_2-\lambda_{345}} , \qquad {\rm or}\qquad \lambda_1=\lambda_2=\lambda_{345}.
\end{align}
For recent studies on the alignment limit in the 2HDM, we refer to ref.~\cite{Carena:2013ooa,Dev:2014yca}.

\begin{table}[t]
\centering
\begin{tabular}{lccccccccc}
\hline
& $\xi_h^u$ & $\xi_h^d$ & $\xi_h^\ell$
& $\xi_H^u$ & $\xi_H^d$ & $\xi_H^\ell$
& $\xi_A^u$ & $\xi_A^d$ & $\xi_A^\ell$ \\ \thickhline
Type-I
& $c_\alpha/s_\beta$ & $+c_\alpha/s_\beta$ & $+c_\alpha/s_\beta$
& $s_\alpha/s_\beta$ & $s_\alpha/s_\beta$ & $s_\alpha/s_\beta$
& $-\cot\beta$ & $+\cot\beta$ & $+\cot\beta$ \\
Type-II
& $c_\alpha/s_\beta$ & $-s_\alpha/c_\beta$ & $-s_\alpha/c_\beta$
& $s_\alpha/s_\beta$ & $c_\alpha/c_\beta$ & $c_\alpha/c_\beta$
& $-\cot\beta$ & $-\tan\beta$ & $-\tan\beta$ \\
Type-X
& $c_\alpha/s_\beta$ & $+c_\alpha/s_\beta$ & $-s_\alpha/c_\beta$
& $s_\alpha/s_\beta$ & $s_\alpha/s_\beta$ & $c_\alpha/c_\beta$
& $-\cot\beta$ & $+\cot\beta$ & $-\tan\beta$ \\
Type-Y
& $c_\alpha/s_\beta$ & $-s_\alpha/c_\beta$ & $+c_\alpha/s_\beta$
& $s_\alpha/s_\beta$ & $c_\alpha/c_\beta$ & $s_\alpha/s_\beta$
& $-\cot\beta$ & $-\tan\beta$ & $+\cot\beta$ \\
\hline
\end{tabular}
\caption{Yukawa  couplings in the four types of 2HDM.}
\label{tab:Yuk}
\end{table}

Since the 2HDMs with $Z_2$ symmetry are particular cases of the A2HDM~\cite{Pich:09}, there exists a one-to-one correspondence for Yukawa couplings between these two models. However, the correspondence is not so straightforward for Higgs cubic couplings. Unlike the 2HDMs with $Z_2$ symmetry, the A2HDM potential is usually defined in the so-called ``Higgs basis''~\cite{Davidson:05}, in which only one Higgs doublet gets a nonzero $vev$. Therefore, the parameter $\tan\beta$ defined in the NFC 2HDMs is not a physical parameter in the A2HDM~\cite{Haber:06}.

\subsection{$B_s \to \mu^+ \mu^-$ in the 2HDMs with $Z_2$ symmetry}
In both the A2HDM and the NFC 2HDMs, $B_s \to \mu^+ \mu^-$ decay is induced by gauge boson $Z$, Goldstone boson $G^0$, and Higgs bosons $\varphi \equiv \lbrace h, H, A \rbrace$ penguin diagrams, as well as box diagrams mediated with $W^\pm$, $H^\pm$, and $G^\pm$. To one-loop level, their contributions to the Wilson coefficients are divided into the following different parts:
\begin{align}\label{eq:WC}
  C_{10}&=\left( C_{10}^{Z,\,\SM} + C_{10}^{\rm box,\,\SM}\right) + \left( C_{10}^{Z,\,\THDM}\right),\\
  C_S&=\left(C_S^{\rm box,\,SM} + C_S^{\rm box,\, 2HDM} + C_S^{\varphi,\,\THDM} \right),\nonumber\\
  C_P&=\left(C_P^{\rm box,\, SM}+C_P^{Z,\, \SM} + C_P^{G,\,\SM}\right) + \left(C_P^{Z,\,\THDM} +C_P^{G,\,\THDM} \right) + \left(C_P^{\rm box,\, 2HDM} + C_P^{\varphi,\,\THDM} \right),\nonumber
\end{align}
where each part in the parentheses is gauge invariant. This gauge invariance is validated by the actual calculation in both the Feynman and the unitary gauges in the A2HDM~\cite{Li:14}. The Wilson coefficients labeled with ``SM'' denote the contributions from the diagrams involved with only the SM fields (with the Goldstone bosons but not the Higgs boson), whose expressions are given in appendix~\ref{sec:app}. Those with ``2HDM'' contain the Higgs contributions. For simplicity, their explicit expressions are given in the unitarity gauge in the following, where the Goldstone boson contributions are absent. 


The Higgs bosons affect the box and $Z$ penguin diagrams with Yukawa interactions. Their contributions to Wilson coefficients in the NFC 2HDMs can easily be obtained from the A2HDM results with replacement of the Yukawa couplings,
\begin{align}\label{eq:repYuk}
  C_{S,P,\,\Unitary}^{\rm box,\, 2HDM}&=C_{S,P,\, \Unitary}^{\rm box,\, A2HDM}\Bigr\rvert_{(\varsigma_u,\varsigma_d,\varsigma_\ell)\to (-\xi_A^u,\xi_A^d,\xi_A^\ell)},\nonumber\\
C_{10,P,\,\Unitary}^{Z,\,\THDM}&=C_{10,P,\, \Unitary}^{Z\,\rm penguin, \, A2HDM} \Bigr\rvert_{(\varsigma_u,\varsigma_d,\varsigma_\ell)\to (-\xi_A^u,\xi_A^d,\xi_A^\ell)}.
\end{align}
For self-contained, we present the Wilson coefficients after the correspondences made in appendix~\ref{sec:app}.
 
The Higgs penguin diagrams involve Yukawa couplings as well as Higgs-gauge couplings and Higgs cubic couplings. Therefore, their Wilson coefficients can not be derived from the A2HDM results so straightforwardly as in the box and $Z$ penguin diagrams, as discussed in previous section. Since the A2HDM Wilson coefficients are given for individual Higgs penguin diagrams in ref.~\cite{Li:14}, we use the following approach.  For every Higgs penguin diagram in the NFC 2HDMs, its contribution is derived from the A2HDM results with the replacement of the Higgs-gauge vertex and the triple Higgs vertex. Then the total contributions to the Wilson coefficients are obtained,
\begin{align}\label{eq:WCHiggsPenguin}
  C_{S,\,\Unitary}^{\varphi,\,\THDM}=&+\frac{x_t\xi_h^\ell}{2x_h}\Bigl(-s_{\alpha-\beta}g_1^{(a)}+c_{\alpha-\beta}g_2^{(a)}+\frac{2v^2}{m_W^2}\lambda_{H^+H^-}^hg_0\Bigr)
\nonumber\\
          &+\frac{x_t\xi_H^\ell}{2x_H}\Bigl(+c_{\alpha-\beta}g_1^{(a)}+s_{\alpha-\beta}g_2^{(a)}+\frac{2v^2}{m_W^2}\lambda_{H^+H^-}^Hg_0\Bigr),
\nonumber\\
  C_{P,\,\Unitary}^{\varphi,\,\THDM}=&-\frac{x_t\xi_A^\ell}{2x_A}g_3^{(a)},
\end{align}
where $x_t = m_t^2/m_W^2$, $x_{h,H,A}=m_{h,H,A}^2/m_W^2$, the functions $g_{0-3}^{(a)}\equiv g_{0-3}^{(a)}\bigl (x_t,x_{H^\pm},-\xi_A^u,\xi_A^d \bigr)$ defined in eq.~\eqref{eq:g}, and  the Higgs cubic couplings are defined as 
\begin{align}\label{eq:cubiccoupling}
\begin{bmatrix}
\lambda^h_{H^+H^-}\\
\lambda^H_{H^+H^-}\\
\lambda^A_{H^+H^-}  
\end{bmatrix}
&=\frac{1}{2v^2s_{2\beta}}
 \begin{bmatrix}
   (m_h^2-2m_{H^\pm}^2)c_{\alpha-3\beta}+(-4M^2+3m_h^2+2m_{H^\pm}^2)c_{\alpha+\beta}\\
   (m_H^2-2m_{H^\pm}^2)s_{\alpha-3\beta}+(-4M^2+3m_H^2+2m_{H^\pm}^2)s_{\alpha+\beta}\\
0
 \end{bmatrix}
,
\end{align}
where the soft $Z_2$ symmetry breaking parameter $M$ has been defined in sec~\ref{sec:2HDM}.

In the literature~\cite{Logan:00,Huang:00,Bobeth:01}, it is found that the Wilson coefficients can receive large $\tan\beta$ enhancement only in the type-II 2HDM and the branching ratio with large $\tan\beta$ depends only on the Higgs masses $m_{H^\pm}$, $m_H$, $m_h$ and the mixing angle $\alpha$. However, as shown by eqs.~\eqref{eq:WCHiggsPenguin} and \eqref{eq:cubiccoupling}, a term proportional to $M^2/m_H^2$ in our full one-loop Wilson coefficient $C_S$ is also enhanced by $\tan^2\beta$, which comes from the heavy Higgs $H$ penguin diagrams mediated by charged Higgs bosons. Using the parameter $m_3$ in the Higgs potential of eq.~\eqref{eq:HiggsPotential} directly, this term is proportional to $m_3^2/m_H^2$ and enhanced by $\tan^3\beta$. This $M$ dependent term has not been considered yet in the previous studies in the literature. Therefore, its effects are worthy of a detailed investigation.

The soft $Z_2$ symmetry breaking parameter $M$ is associated with the spontaneous CP breaking~\cite{Lee:73,Ivanov:06,Ferreira:09,Grzadkowski:14} and characterizes the masses of all the Higgs bosons~\cite{Kanemura:04}. This parameter enters the $B_s \to \mu^+ \mu^-$ decays through the Higgs penguin diagrams. However, it is found that the $M$ term can not make more significant contributions than other terms of the Wilson coefficient $C_S$. Here, we would choose $h$ as the Higgs boson discovered by ATLAS~\cite{ATLAS:12} and CMS~\cite{CMS:12} and take the alignment limit $\beta-\alpha=\pi/2$, which is favored by the current 2HDM fits~\cite{Coleppa:13,Chen:13,Eberhardt:13,Dumont:14a,Dumont:14b,Bernon:14,Chowdhury:15}. Then the cubic couplings in eq.~\eqref{eq:cubiccoupling} read
\begin{align}\label{eq:cubiccoupling:2}
\begin{bmatrix}
\lambda^h_{H^+H^-}\\
\lambda^H_{H^+H^-}\\
\lambda^A_{H^+H^-}  
\end{bmatrix}
\doteq\frac{1}{v^2}
\begin{bmatrix}
  -2M^2+2m_{H^\pm}^2+m_h^2\\
  \cot{2\beta} (2M^2-2m_H^2)\\
0
\end{bmatrix}.
\end{align}
Focusing on the coupling $\lambda_{H^+ H^-}^h$, it can be seen from eqs.~\eqref{eq:SP} and \eqref{eq:cubiccoupling:2} that large contributions from this coupling would require $\lvert M^2-m_{H^\pm}^2 \rvert/v^2\gg m_W^2/m_B^2$. However, we know $\lvert M^2-m_{H^\pm}^2 \rvert / v^2=\lvert \lambda_4+\lambda_5 \rvert /2 < 4\pi$ from the 2HDM vacuum condition~\cite{Kanemura:04} and perturbativity~\cite{Kanemura:99}. It is also noted that, the Higgs penguin diagrams can be enhanced by very large $\tan\beta$ or $\cot\beta$. In all the four types of 2HDMs, the $\lambda_{H^+H^-}^h$ contributions could be enhanced by large $\cot^2\beta$. In practice, $\cot\beta \gtrsim 3$ has been excluded by the perturbativity~\cite{Kanemura:99}. Similarly, the coupling $\lambda_{H^+H^-}^H$ can make a large contribution if $M^2/m_H^2 \gg m_W^2/m_B^2$. Among the four models, this contribution is enhanced by $\tan^2\beta$ only in type-II 2HDM. However, the ratio $M^2/m_H^2$ still suffers from the theoretical constraints, which will be discussed with numerical results in the following section.

Although the effects from the operators $\mathcal O_S$ and $\mathcal O_P$ are suppressed by $m_B^2/m_W^2$, these two scalar operators can make significant contributions in the two parameter regions: (i) in the type-II 2HDM, since both $C_S$ and $C_P$ contain $\tan\beta$ enhanced terms, the effects of the scalar operators are enhanced in the parameter space with large $\tan\beta$. (ii) The contributions from the CP-odd Higgs penguin diagrams are inversely proportional to the mass of the CP-odd Higgs boson $A$. Thus, the Wilson coefficient $C_P$ becomes much more significant in the region with small values of $m_A$\footnote{For the CP-odd Higgs in the MSSM, the LEP experiment put a lower bound on its mass $m_A>93.4\GeV$~\cite{LEP:06}.} in all the four 2HDMs.

In the particular case of the type-II 2HDM, our result of $C_{10}$ agrees with the one calculated in ref.~\cite{Chankowski:00}. For the Wilson coefficients $C_S$ and $C_P$ in the 2HDM, the calculations have been performed by various groups~\cite{He:88,Savage:91,Skiba:92,Dai:96,Huang:98,Choudhury:98,Logan:00,Huang:00,Bobeth:01}. The latest results are presented in these three papers~\cite{Logan:00,Huang:00,Bobeth:01}, where the 2HDM contributions are computed in the type-II model in some specific cases. In ref.~\cite{Logan:00}, the Wilson coefficients are calculated in large $\tan\beta$ limit, i.e., only $\tan^2\beta$ enhanced terms are kept. However, the Higgs penguin diagrams with trilinear $hH^+ H^-$ and $HH^+ H^-$ couplings are not considered. In refs.~\cite{Huang:00} and \cite{Bobeth:01}\footnote{In ref.~\cite{Bobeth:01}, it is mentioned that their result is different from the one in ref.~\cite{Huang:00}. However, the two results agree with each other after the erratum for ref.~\cite{Huang:00} has been taken into account. In addition, there is a typo in eqs.~(3.30) and (3.31) of ref.~\cite{Huang:00}: a global factor $\alpha_e/\pi$ should be included.}, after including these penguin diagrams, the calculations are performed  again in the large $\tan\beta$ limit but with the assumption $\lambda_1 = \lambda_2$ for the couplings in the Higgs potential\footnote{In ref.~\cite{Huang:00,Bobeth:01}, the convention for the Higgs potential (i.e., the couplings $\lambda_i$) is different from the one defined in eq.~\eqref{eq:HiggsPotential}. This condition is also expressed as $\lambda_1=\lambda_2$ by the couplings used in our paper.}. Considering only terms proportional to $\tan^2\beta$, our result agrees with the one of ref.~\cite{Logan:00} in the case of $\lambda_{H^+H^-}^h=\lambda_{H^+H^-}^H=0$, and those of ref.~\cite{Huang:00,Bobeth:01} in the case of $\lambda_1=\lambda_2$. Generally, the 2HDM contains eight free parameters, i.e., $m_{1-3}$ and $\lambda_{1-5}$ in the Higgs potential of eq.~\eqref{eq:HiggsPotential}. They can be rewritten equivalently in terms of the Higgs masses $m_h$, $m_H$, $m_A$, $m_{H^\pm}$, the mixing angles $\alpha$ and $\beta$, the parameter $M$, and the $vev$ $v=v_{\rm SM}$. If the condition $\lambda_1=\lambda_2$ is assumed, $M$ can be expressed by the other parameters, as shown in eq.~\eqref{eq:lam:2}. It is the reason why terms depending on the $Z_2$ symmetry breaking parameter $M$ were absent in the previous calculations~\cite{Logan:00,Huang:00,Bobeth:01}, but are present in this paper.

\section{Numerical Analysis}\label{sec:analysis}

Searches for $B_{s,d} \to \mu^+ \mu^-$ decays have been performed at the BaBar, Belle, and Tevatron (for a review, see ref.~\cite{Albrecht:12}). At the LHC, measurements by CMS~\cite{CMS:13a} and LHCb~\cite{LHCb:13} collaborations with the full data of LHC Run I have resulted in the averaged values for the time-integrated branching ratios~\cite{LHC:2014}
\begin{align*}
\B (B_s \to \mu^+ \mu^-) &= 2.8_{-0.6}^{+0.7} \times 10^{-9},\\
\B (B_d \to \mu^+ \mu^-) &= 3.9_{-1.4}^{+1.6} \times 10^{-10},
\end{align*}
where the errors are dominated by the statistical uncertainties and expected to be significantly reduced in the near future. Both of them are in good agreement with the latest updated SM predictions~\cite{Bobeth:13a}, $\B (B_s \to \mu^+ \mu^-)=(3.65\pm 0.23)\times 10^{-9}$ and $\B (B_d \to \mu^+ \mu^-)=(1.06\pm 0.09)\times 10^{-10}$, in which the NLO EW~\cite{Bobeth:13b} and the NNLO QCD~\cite{Hermann:13} corrections have been included. Thus, strong constraints on the 2HDM parameters are expected.

In the NFC 2HDMs, the relevant parameters are the two mixing angles $\alpha$ and $\beta$, four Higgs mass parameters $m_{H^\pm}$, $m_h$, $m_H$, and $m_A$. In the $B_{s,d} \to \mu^+ \mu^-$ decays, the $Z_2$ symmetry breaking parameter $M$ also enters into the decay amplitude and is independent from these parameters. As discussed in ref.~\cite{Carmi:12,Craig:13}, we choose the light neutral Higgs $h$ in the 2HDM as the SM Higgs observed  at the LHC and adopt the alignment limit $\sin(\beta-\alpha)=1$. Then the model parameters are reduced to $(m_H,m_A,m_{H^\pm},M,\tan\beta)$. As discussed in ref.~\cite{Cheng:14}, we shall restrict these parameters in the following ranges:
\begin{align}\label{eq:PS}
  m_H \in [m_h,1000]\GeV, \qquad m_{H^\pm},m_A,M\in [1,1000]\GeV,\qquad \tan\beta \in [0.1, 100].
\end{align}
Starting from these parameter spaces, we will start our numerical scan.

In the numerical analysis, we impose experimental constraints in the same way as in ref.~\cite{Cheng:14}. To constrain the 2HDM parameters, we 
have taken into account  (i) flavor processes: $B_{s,d}-\bar B_{s,d}$ mixing, $\bar B \to X_s \gamma$, $B \to \tau \nu$ and $B_{s,d}\to \mu^+ \mu^-$ decays, (ii) direct searches for Higgs bosons at LEP~\cite{ALEPH:02}, Tevatron~\cite{CDF:05,D0:09} and LHC~\cite{CMS:12a,ATLAS:13a}, both of which have been discussed in detail in our previous work~\cite{Cheng:14}. Additionally, we also consider the oblique parameter $\Delta\rho$ in the EW precision measurement~\cite{Peskin:90,Altarelli:90,Peskin:91,Altarelli:91,Grimus:07,Grimus:08} and require the couplings $\lambda_{1-5}$ to satisfy (iii) theoretical constraints: perturbativity~\cite{Kanemura:99}, tree-level vacuum stability~\cite{Deshpande:77,Nie:98,Ivanov:06} and perturbative unitarity~\cite{Arhrib:00,Branco:11} (See ref.~\cite{Chang:15} for the expressions).

\begin{figure}[t]
  \centering
  \subfigure{\label{fig:M}\includegraphics[width=0.45\textwidth]{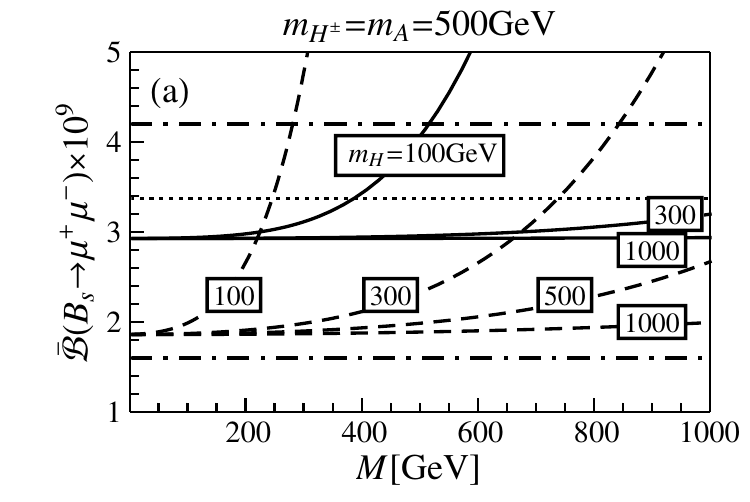}}
  \quad
  \subfigure{\label{fig:PS}\includegraphics[width=0.45\textwidth]{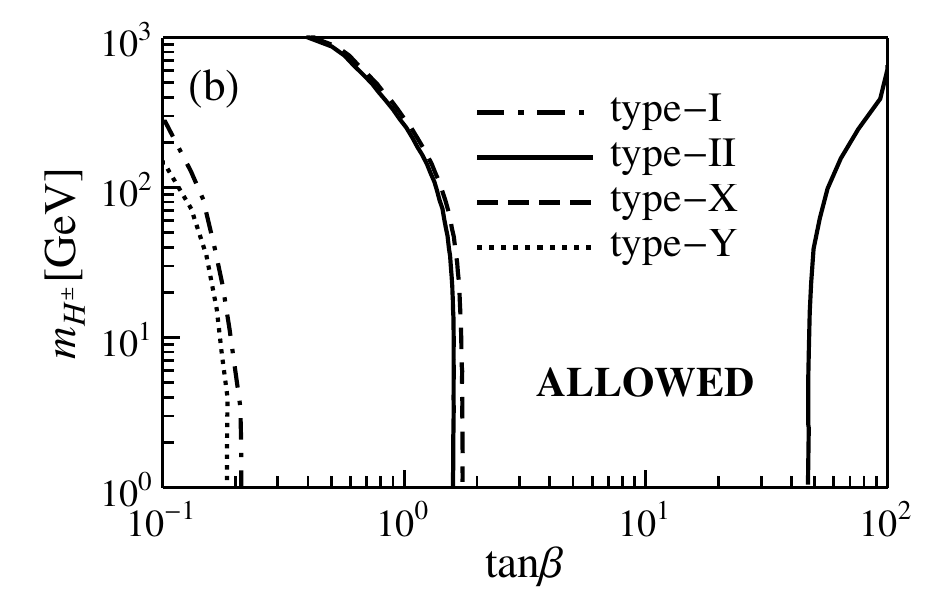}}
  \caption{(a) The $M$ dependence of the branching ratio of $B_s \to \mu^+ \mu^-$ in the type-II 2HDM for $\tan\beta=20$ (solid) and $\tan\beta=40$ (dashed). The SM prediction (dotted) and $2\sigma$ experimental range (dot-dashed) are also shown. (b) Allowed regions of the parameter space $(\tan\beta,m_{H^\pm})$ from $\overline{\mathcal B} (B_s \to \mu^+ \mu^-)$ for the four types of 2HDM.}
\end{figure}

For $B_s \to \mu^+ \mu^-$ decay, both the NNLO QCD and the NLO EW corrections in the SM and the full one-loop contributions in the 2HDM are included. As discussed in sec.~\ref{sec:WC}, the effects of the soft $Z_2$ symmetry breaking parameter $M$ can be enhanced by large $\tan\beta$ in the type-II 2HDM. The $M$ dependence of the branching ratio $\B (B_s \to \mu^+ \mu^-)$ is shown in figure~\ref{fig:M} in the type-II 2HDM for various $\tan\beta$ and $m_H$ values. As expected, the effects of $M$ become significant when the two ratios $M^2/m_H^2$ and $\tan\beta$ are large. However, it is found that the theoretical constraints from perturbativity, vacuum stability, and perturbative unitarity have put the bound $M^2/m_H^2 \lesssim 1$ (and $M \lesssim 1\TeV$) in the parameter space of eq.~\eqref{eq:PS}. Therefore, the soft $Z_2$ symmetry breaking parameter $M$ can not make more significant effects than the other $\tan\beta$ enhanced terms in $C_S$ and $C_P$.


After considering the current experimental data, the allowed parameter spaces of all the four 2HDMs are obtained. Since the constraints from $B_d \to \mu^+ \mu^-$ appear to be more or less weaker than those from $B_s \to \mu^+ \mu^-$, we only show the results from the latter one, which are plotted in the $(\tan\beta,m_{H^\pm})$ plane in figure~\ref{fig:PS}. Compared to our previous results~\cite{Cheng:14}, the parameter space with small $\tan\beta$ is excluded for all the four types of 2HDMs. This change is caused by the contributions with small $\tan\beta$ neglected in the previous calculations~\cite{Logan:00,Huang:00,Bobeth:01} but included in the present full one-loop computation as discussed in section.~\ref{sec:WC}. 
For the large $\tan\beta$ region, only the type-II model is bounded, which is in agreement with our previous result but still weaker than the one from $\mathcal B (B \to \tau\nu)$.

\begin{figure}[t]
  \centering
  \subfigure{\label{fig:PS_All}\includegraphics[width=0.45\textwidth]{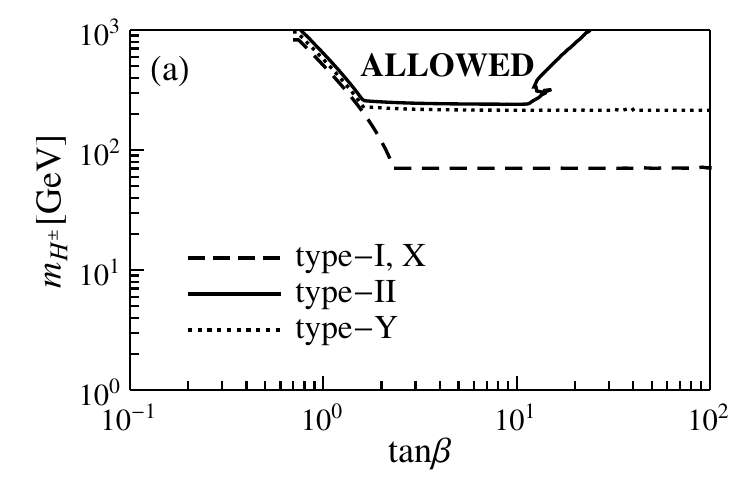}}
  \quad
  \subfigure{\label{fig:correlation}\includegraphics[width=0.45\textwidth]{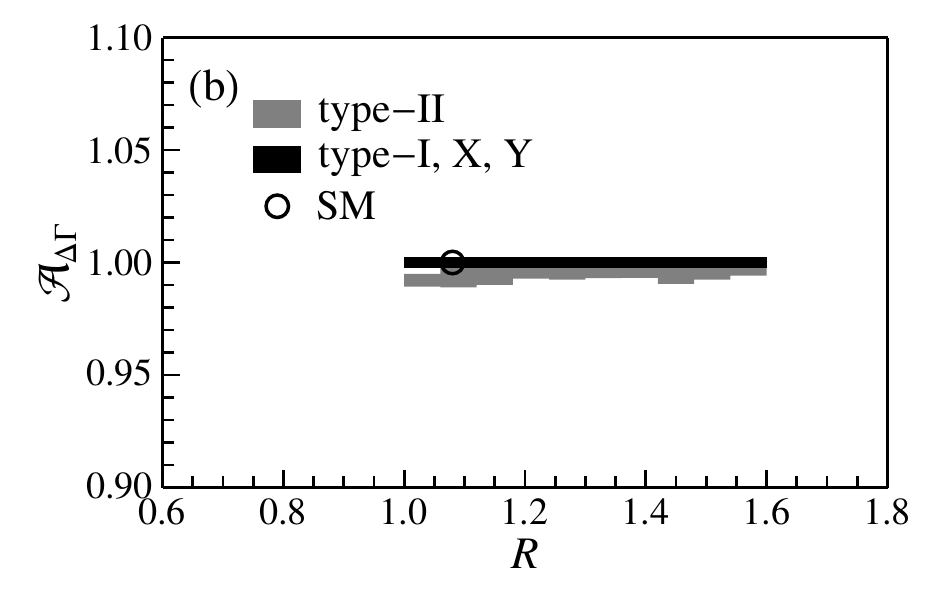}}\\
  \caption{(a) Combined constraints on the parameter space of the four types of 2HDM, plotted in the $(\tan\beta,m_{H^\pm})$ plane.  (b) Correlations between $R$ and $\mathcal A_{\Delta\Gamma}$ in the four types of 2HDM.}
\end{figure}

Combining all the constraints aforementioned, we obtain the survived parameter space of all the four types 2HDMs, as an update of our previous results~\cite{Cheng:14}, which is shown in the $(\tan\beta, m_{H^\pm})$ plane in figure~\ref{fig:PS_All}. It is found that the small $\tan\beta$ region is restricted for all the four models by $B_s - \bar B_s$ mixing and $B \to X_s \gamma$ , while the large $\tan\beta$ region is constrained only in the type-II 2HDM by $B\to \tau \nu$ and $B_s \to \mu^+ \mu^-$ decays. 
Compared to our previous results, the current constraints on the large $\tan\beta$ region in the type-II 2HDM are more stringent. This is mainly because the theoretical constraints are included in the current analysis.


In these  constrained parameter spaces of the four 2HDMs, the correlations between the observables $\mathcal A_{\Delta\Gamma}$ and $R$ defined in eqs.~\eqref{eq:A} and \eqref{eq:R} are reevaluated, which are presented in figure~\ref{fig:correlation}. Unlike our previous results~\cite{Cheng:14}, the correlations in the four different types of 2HDMs are almost indistinguishable. The allowed ranges of $R$ are the same for all the four models, while $\mathcal A_{\Delta\Gamma}$ can deviate slightly from the SM prediction only in the type-II 2HDM. 
It is found that the difference from our previous results is mainly caused by the theoretical constraints and the new full one-loop Wilson coefficients considered in the current analysis. In the type-II 2HDM, the bounds on $\tan\beta$ are more stringent compared to our previous results as discussed above. Thus, the allowed range of $C_S$ is restricted more stringently in the current analysis. In this case, $\mathcal A_{\Delta\Gamma}$ can  deviate from the SM prediction very tiny, which can be seen from eq.~\eqref{eq:A}. As discussed in sec.~\ref{sec:WC}, our results of the Wilson coefficients can also be applied to the small $\tan\beta$ region in all the four models, while some terms are not included in $C_P$ used in our previous analysis. In the case of small $m_A$, $C_P$ is enhanced and these terms make the allowed regions of $R$ in the type-I and type-Y 2HDMs as large as the one in the type-X model. Meanwhile, the value of $R$ is almost independent of $\mathcal A_{\Delta\Gamma}$ in the type-II 2HDM.

\section{Conclusion}\label{sec:conclusion}
In this paper, we have performed an updated analysis of the rare leptonic decay $B_s \to \mu^+ \mu^-$ in the 2HDM with a softly broken $Z_2$ symmetry. We have derived the full one-loop Wilson coefficients $C_{10}$, $C_S$ and $C_P$ from the recent A2HDM results~\cite{Li:14}, which can be applied to the contributions of all the four types of 2HDMs for both large and small $\tan\beta$ value. Our main conclusions are summarized as follows:

\begin{itemize}
\item Compared to $C_{10}$, the Wilson coefficients $C_S$ and $C_P$ are negligible in the entire 2HDM parameter space, except for large $\tan\beta$ in the type-II 2HDM or small CP-odd Higgs mass $m_A$ in the four models. In addition, only the Wilson coefficients $C_S$ and $C_P$ in the type-II 2HDM can be enhanced by large $\tan\beta$.
\item The soft $Z_2$ symmetry breaking parameter $M$ enters into the Higgs penguin diagrams and affects the Wilson coefficient $C_S$. The dominant contributions are proportional to $M^2/m_H^2$ and enhanced by $\tan^2\beta$ in the type-II 2HDM, which have not been considered in the literature~\cite{Logan:00,Huang:00,Bobeth:01}. However, after combing the theoretical constraints from perturbativity, vacuum stability and perturbative unitarity, we have found that the parameter $M$ can not make more significant contributions than other terms in the Wilson coefficients.
\item After imposing the experimental constraints, regions with small $\tan\beta$ are excluded for all the four types of 2HDM, which are quite different from our previous results~\cite{Cheng:14}. As expected, large $\tan\beta$ region is only excluded in the type-II 2HDM.
\end{itemize}

As an update of our previous analysis~\cite{Cheng:14}, we have also investigated the possibility to distinguish the four types of 2HDM in light of the recent  updated flavor physics data, the collider data from the direct searches for Higgs bosons and the theoretical progresses. The combined bounds on the 2HDM parameters have been derived for the four models. In the survived parameter regions, the correlations between $\mathcal A_{\Delta\Gamma}$ and $R$ in all the four of the 2HDMs are almost indistinguishable from each other.  In the 2HDMs with $Z_2$ symmetry, $\mathcal A_{\Delta\Gamma}$ can only have a very tiny deviation from the SM prediction, while  $R$ could  deviate from the SM one sizably. This could be tested by the much more precise measurement of $B_s \to \mu^+ \mu^-$ at the LHC in the coming years.

\section*{Acknowledgments}
The work was supported by the National Natural Science Foundation of China (NSFC) under contract Nos.~11225523, 11221504 and 11435003. XY is also supported in part by the NRF grant funded by the Korean government of the MEST (No.~2011-0017430) and (No.~2011-0020333), and by the Open Project Program of SKLTP, ITP, CAS (No.~Y4KF081CJ1). XY acknowledges the hospitality of the ITP, CAS, where this work was finalized. We thank Xin-Qiang Li for useful discussions.

\begin{appendix}
\section{The Wilson coefficients in the SM and the 2HDMs}\label{sec:app}
In this appendix, we recapitulate the relevant expressions of the Wilson coefficients in the SM and the four types of the 2HDMs for completeness, which are obtained from ref.~\cite{Li:14}.

In the SM, the one-loop Wilson coefficients of the scalar operators can be written as
\begin{align}\label{eq:WCSPSM}
  C_S^\SM&=C_S^{\rm box,\,SM}+C_S^{h,\,\SM}\,,\\
  C_P^\SM&=C_P^{\rm box,\,\SM}+C_P^{Z,\,\SM}+C_P^{G,\,\SM}\,.\nonumber
\end{align}
In the unitary gauge, their expressions read
\begin{align}\label{eq:WCSM}
  C_{S,\,\Unitary}^{\rm box,\,SM}=&-\frac{x_t(x_t+1)}{48(x_t-1)^2} - \frac{(x_t-2)(3x_t^2-3x_t+1)}{24(x_t-1)^3}\, \ln x_t\, ,
\\
  C_{S,\,\Unitary}^{h,\,\SM}=&-\frac{3 \xt}{8 x_h}\,,
\nonumber\\
  C_{P,\,\Unitary}^{\rm box,\,\SM}=&+\frac{x_t(71x_t^2-172 x_t-19)}{144(x_t-1)^3} + \frac{x_t^4-12x_t^3+34 x_t^2-x_t-2}{24(x_t-1)^4} \ln x_t \,,
\nonumber\\
  C_{P,\,\Unitary}^{Z,\,\SM}=&+\frac{1}{12}\left[\frac{x_t(18 x_t^3-137 x_t^2+262 x_t-95)}{6 (x_t-1)^3} + \frac{8 x_t^4-11 x_t^3-15 x_t^2+12 x_t-2}{(x_t-1)^4}\ln x_t \right]\nonumber \\
&- \frac{s_W^2}{36}\left[\frac{x_t(18 x_t^3-139 x_t^2+274 x_t-129)}{2 (x_t-1)^3} + \frac{24 x_t^4-33 x_t^3-45 x_t^2+50 x_t-8}{(x_t-1)^4} \ln x_t\right]\,, \nonumber
\end{align}
where $C_{S,\,\Unitary}^{h,\,\SM}$ denotes the contributions from the SM Higgs penguin diagrams. The other Wilson coefficients $C_{S,P,\,\Unitary}^{\rm box,\,\SM}$ and $C_{P,\,\Unitary}^{Z,\,\SM}$ are same in the SM and the 2HDMs.

In the four types of the 2HDMs, the various contributions in the Wilson coefficients of eq.~\eqref{eq:WC} are obtained by the replacement of the Yukawa couplings in eq.~\eqref{eq:repYuk}, which are given in the unitary gauge,
\begin{align}
  C^{\rm box,\,\THDM}_{S,\,\rm Unitary}=&
  -\frac{\xi_A^u\xi_A^\ell\xt}{8(\xHC-\xt)} \left[1-\frac{\xHC \ln(\xHC/\xt)}{(\xHC-\xt)}\right]
  -\xi_A^d\xi_A^\ell \frac{\xt\ln(\xHC/\xt)}{4(\xHC-\xt)}\,,
\\[0.8em]
  C_{P,\,\Unitary}^{\rm box,\,\THDM}=&
  +\frac{\xi_A^u\xi_A^\ell\xt}{8(\xHC-\xt)} \left[1 +\frac{2\xt^2-\xHC\xt-\xHC}{(\xt-1)(\xHC-\xt)}\ln\xt + \frac{\xHC (1-2\xt +\xHC)}{(\xHC-1)(\xHC-\xt)}\ln\xHC \right]
\nonumber\\
  &+\xi_A^d\xi_A^\ell \frac{\xt\ln(\xHC/\xt)}{4(\xHC-\xt)}\,,
\nonumber\\[0.8em]
  C^{Z,\, \THDM}_{10,\,\Unitary}=& + (\xi_A^u)^2\,\frac{x_t^2}{8}\left[\frac{1}{\xHC-x_t} - \frac{\xHC\ln(\xHC/\xt)}{(\xHC-x_t)^2}\right]\,,&
\nonumber\\[0.8em]
  C_{P,\,\Unitary}^{Z,\,\THDM}=&
  +\frac{x_t}{4(\xHC-x_t)^2}\, \biggl\lbrace -\xi_A^d\xi_A^u\biggl[- \frac{x_t+\xHC}{2} + \frac{x_t \xHC}{\xHC-x_t}\ln\frac{\xHC}{\xt}\biggr]
 \nonumber\\
 &\hspace{7.38em} + (\xi_A^u)^2\biggl[\frac{\xHC^2-8\xHC x_t-17x_t^2}{36(\xHC-x_t)} -\xt (\xHC-\xt)
 \nonumber\\
 &\hspace{11.3em}+\left(\frac{x_t^2(3\xHC+x_t)}{6(\xHC-x_t)^2}+\xt\xHC\right)\ln\frac{\xHC}{\xt}\biggr]\biggr\rbrace 
\nonumber\\[0.5em]
&+ \frac{s_W^2 x_t}{6(\xHC-x_t)^2}\,\biggl\lbrace-\xi_A^d\xi_A^u\,\biggl[\frac{5 x_t-3 \xHC}{2} + \frac{\xHC(2\xHC-3x_t)}{\xHC-x_t}\ln\frac{\xHC}{\xt}\biggr] 
\nonumber\\
&\hspace{7.52em} - (\xi_A^u)^2\biggl[\left(\frac{4 \xHC^3-12 \xHC^2 x_t+9 \xHC x_t^2+3 x_t^3}{6(\xHC-x_t)^2}+\frac{3}{2}\xt\xHC\right)\ln\frac{\xHC}{\xt} 
\nonumber\\
 &\hspace{11.3em}- \frac{17 \xHC^2-64 \xHC x_t+71 x_t^2}{36(\xHC-x_t)} -\frac{3}{2}\xt(\xHC-\xt)\biggr]\biggr\rbrace \,.\nonumber
\end{align}

The Higgs penguin contributions $C_{S,P,\,\Unitary}^{\varphi,\,\THDM}$ have been given in eq.~\eqref{eq:WCHiggsPenguin}, where the functions $g_{0-3}^{(a)}$ are defined as 
\begin{align}\label{eq:g}
g_0(x_t,\xHC,-\xi_A^u,\xi_A^d) &=-\frac{1}{4\xHC}\left[-\xi_A^u\xi_A^d(f_1+f_2+f_3+1)+(\xi_A^u)^2\left(f_4-f_5-\frac{1}{4}\right)\right]\,,
\\
g_1^{(a)}(x_t,\xHC,-\xi_A^u,\xi_A^d) &=
    -\frac{\,3\,}{4} -\xi_A^u\xi_A^d (f_1+f_2+f_3) + (\xi_A^u)^2 (f_4-f_5)\,,
\nonumber\\
g_2^{(a)}(x_t,\xHC,-\xi_A^u,\xi_A^d) &=
   -(\xi_A^d)^2\xi_A^u  f_1
   +\xi_A^d(\xi_A^u)^2 ( f_3 + f_2) 
   +(\xi_A^u)^3 (f_5 - f_4)
   +\xi_A^u (f_7 -  f_6)
   +\xi_A^d f_1\,,
\nonumber\\
g_3^{(a)}(x_t,\xHC,-\xi_A^u,\xi_A^d) &=
   -(\xi_A^d)^2 \xi_A^u  f_1
   +\xi_A^d(\xi_A^u)^2 (f_3 - f_2) 
   -(\xi_A^u)^3 ( f_5 + f_4)
   -\xi_A^u ( f_7 + f_6 )
   +\xi_A^d f_1\,.\nonumber
\end{align}
Here the one-loop functions $f_i$ are abbreviated as $f_i\equiv f_i(x_t,x_{H^\pm})$ with the definitions
\begin{align}
 f_1(x,y ) =& \frac{1}{2(y- x)^{\phantom{2}}}\,\biggl[x-y +y \ln y-x \ln x\biggr]\,, 
\\
 f_2(x,y)  =& \frac{1}{2(y-x)^{\phantom{2}}} \left[x - \frac{y x}{y-x}(\ln y-\ln x)\right]\,, 
\nonumber\\
 f_3(x,y)  =&\frac{1}{2(y-x)^{\phantom{2}}} \left[y-\frac{y^2 \ln y}{y-x}+\frac{x (2y- x) \ln x}{y-x}\right]\,, 
\nonumber\\
 f_4(x,y)  =&\frac{1}{4(y-x)^2}
     \left[\frac{x \left(3 y-x\right)}{2}-\frac{y^2 x}{y- x}(\ln y-\ln   x)\right]\,, 
\nonumber\\
 f_5(x,y) =&\frac{1}{4(y-x)^2}
   \left[\frac{x (y-3 x)}{2}-\frac{y x (y-2 x)}{y- x}(\ln y-\ln   x)\right]\,, 
\nonumber\\
 f_6(x,y) =&\frac{x\left(x^2-3yx+9y-5 x-2\right)}{8(x-1)^2(y-x)}
  +\frac{y \left(y x-3y+2 x\right) }{4 (y-1)(y- x)^2}\ln y\nonumber \\[0.2cm]
  &+\frac{y^2 \left(-2 x^3+6 x^2-9
   x+2\right)+3 y x^2 (x^2-2 x+3)-x^2
   \left(2 x^3-3 x^2+3 x+1\right)}{4 (x-1)^3 (y- x)^2}\ln x\,, 
\nonumber\\
 f_7(x,y) =&\frac{\left(x^2+x-8\right)}{8 (x-1)^2}-\frac{y   (y+2) }{4 (y-1)(y-x)}\ln y+\frac{y \left(x^3-3 x^2+3 x+2\right)+3 \left(x-2\right) x^2}{4 (x-1)^3(y-x)}\ln x\,.\nonumber
\end{align}
It is noted that the divergence in the Higgs penguin diagrams at one-loop level is canceled by a FCNC local operator in the A2HDM~\cite{Li:14}. In the 2HDMs with $Z_2$ symmetry,  we find that the divergence automatically vanishes after adding all the Higgs penguin contributions.

For the four types of 2HDM, the values of the relevant Yukawa couplings are listed in table~\ref{tab:Yuk}. When deriving the expressions of the Higgs penguin diagrams in eq.~\eqref{eq:WCHiggsPenguin}, the following identities have been used:
\begin{align}\label{eq:identity:1}
  \xi_h^u&=-\cos(\alpha-\beta) \xi_A^u - \sin(\alpha-\beta),&
  \xi_H^u&=-\sin(\alpha-\beta) \xi_A^u + \cos(\alpha-\beta),\\
  \xi_h^d&=+\cos(\alpha-\beta) \xi_A^d - \sin(\alpha-\beta),&
  \xi_H^d&=+\sin(\alpha-\beta) \xi_A^d + \cos(\alpha-\beta),\nonumber
\end{align}
and
\begin{align}\label{eq:identity:2}
  (\xi_A^u+\xi_A^d)(\xi_A^u\xi_A^d-1)=0,
\end{align}
which can be obtained from table~\ref{tab:Yuk}. It should be noted that there is a freedom in the definitions of the functions $f_i$, since adding the LHS of eq.~\eqref{eq:identity:2} to eq.~\eqref{eq:g} does not change $g_{0-3}^{(a)}$.

\end{appendix}

\end{document}